\documentclass[twocolumn,superscriptaddress,nofootinbib,amsmath,amssymb, amsfonts]{aastex631}

\usepackage{times}
\usepackage{amsmath}
\usepackage{mathtools}

\newcommand{\cf}{cf.,~}
\newcommand{\ie}{i.e.,~}
\newcommand{\eg}{e.g.,~}

\shorttitle{Post-merger gravitational-wave signal: A new look}
\shortauthors{Topolski, Tootle and Rezzolla}

\graphicspath{{figs/}}

\begin{document}

\title{{Post-merger} gravitational-wave signal from neutron-star binaries: a new look
  at an old problem}

\author[0000-0001-9972-7143]{Konrad Topolski}
\affiliation{Institut f\"ur Theoretische Physik, Goethe Universit\"at,
Max-von-Laue-Str. 1,
60438 Frankfurt am Main, Germany}

\author[0000-0001-9781-0496]{Samuel D. Tootle}
\affiliation{Institut f\"ur Theoretische Physik, Goethe Universit\"at,
Max-von-Laue-Str. 1,
60438 Frankfurt am Main, Germany}

\author[0000-0002-1330-7103]{Luciano Rezzolla}
\affiliation{Institut f\"ur Theoretische Physik, Goethe Universit\"at,
Max-von-Laue-Str. 1,
60438 Frankfurt am Main, Germany}
\affiliation{Frankfurt Institute for Advanced Studies,
Ruth-Moufang-Str. 1,
60438 Frankfurt am Main, Germany}
\affiliation{School of Mathematics, Trinity College,
Dublin 2, Ireland}

 
\begin{abstract}
The spectral properties of the {post-merger} gravitational-wave signal
from a binary of neutron stars encodes a variety of information about the
features of the system and of the equation of state describing matter
around and above nuclear saturation density. Characterising the
{properties} of such a signal is an ``old'' problem, which first emerged
when a number of frequencies were shown to be related to the properties
of the binary through ``quasi-universal'' relations. Here we take a new
look at this old problem by computing the properties of the signal in
terms of the Weyl scalar $\psi_4$. In this way, and using a database of more than
100 simulations, we provide the first evidence for a new {instantaneous
  frequency}, $f^{\psi_4}_0$, associated with the instant of quasi
time-symmetry in the postmerger dynamics, and which also 
follows a quasi-universal relation. We also derive a
new quasi-universal relation for the merger frequency $f^{h}_{\rm mer}$, 
which provides a description of the data that is four times more accurate 
than previous expressions while requiring fewer fitting coefficients. 
Finally, {consistently with the findings of numerous studies before ours,
 and using an enlarged  ensamble of binary systems} we point out that the $\ell=2, m=1$
gravitational-wave mode could become comparable with the traditional
$\ell=2, m=2$ mode on sufficiently long timescales, with strain
amplitudes in a ratio $|h^{21}|/|h^{22}| \sim 0.1-1$ under generic
orientations of the binary, which could be measured by present detectors
for signals with large signal-to-noise ratio or 
by third-generation detectors for generic signals should no collapse occur.
\end{abstract}


\section{Introduction}
\label{sec:intro}

The observation of a gravitational-wave (GW) signal from the binary
neutron-star (BNS) merger event GW170817~\citep{Abbott2017}, and the
detection of an electromagnetic (EM) counterpart, has testified the
enormous potential of GW astronomy. {Starting from early works with
  simplified equations of state (EOSs)~\citep[see, \eg][]{Shibata05c,
    Anderson2007, Liu:2008xy, Baiotti08, Hotokezaka2011}, increasingly}
more comprehensive simulations of these events, which involve an ever
more detailed description of the microphysics~\citep{Bauswein:2018bma,
  DePietri2019, Gieg2019, Tootle2022, Most2022, Camilletti2022,
  Ujevic2023}, of the magnetic-field evolution~\citep{Rezzolla:2011,
  Dionysopoulou:2012, Ciolfi2019, Sun:2022vri, Zappa2023}, and its
amplification~\citep{Kiuchi2015a, Palenzuela_2022b, Chabanov2022}, and of
transport of neutrinos~\citep{Foucart2022aa, Zappa2023}, allow one to
make predictions from the early inspiral up to the long-term evolution of
the postmerger remnant~\citep{DePietri2020, Kiuchi2022}. During each
stage in the evolution of the binary, the features of the GW and EM
signals change in a characteristic manner, encoding information on the
properties of the constituent neutron stars and of the hypermassive
neutron star (HMNS) produced after the merger and, hence, on the
governing EOS.

Characterising the {properties of the post-merger} GW signal is a rather
``old'' problem, which has first emerged when a number of peculiar
frequencies were shown to be related with the properties of the binary
through \textit{quasi-universal} relations, \ie relations that are almost
independent of the specific EOS. These relations have been suggested for
the GW frequency at merger $f_{\rm mer}$ \citep{Read2013, Bernuzzi2014,
  Takami2015, Rezzolla2016, Most2018b, Bauswein2019, Weih:2019xvw,
  Gonzalez2022}, the dominant frequency in the postmerger spectrum
$f_{2}$ {\citep[see, \eg][]{Oechslin07b, Bauswein2011, Read2013,
    Rezzolla2016, Gonzalez2022}}, and other frequencies identifiable in
the transient period right after the merger~\citep{Bauswein2015,
  Takami2015, Rezzolla2016}. Fits to these quasi-universal relations have
been employed in a number of studies (see, \eg~\citep{Bauswein2015b,
  Baiotti2016} for some reviews). These EOS-insensitive relations can
help enormously in constraining the EOS of matter at nuclear densities,
marking the possible appearance of phase transitions~\citep{Most:2018eaw,
  Weih2020, Liebling2021, Prakash:2021wpz, Fujimoto:2022c, Tootle2022,
  Espino2023}, inform waveform models~\citep{Bose2017, Breschi:2019srl};
{however, see~\citet{Raithel2022} for possible violations of these
  relations.}

Another relatively ``old'' problem in the characterisation of the GW
signal from BNS is the one about the relative weight of the lower-order
multipole $\ell=2,m=1$. Numerical simulations have highlighted that the
HMNS can be subject to a nonaxisymmetric instability that powers the
growth of $\ell=2,m=1$ mode of the rest-mass density distribution and,
hence, of the corresponding GW signal~{\citep[see, \eg][]{East2015,
    Lehner2016, Lehner2016a, Radice2016a, East2019,
    Papenfort:2022ywx}}. This mode can already be seeded by the initial
asymmetry of the system in the unequal-mass case, or develop by the
shearing of the contact layers of the binary constituents upon
merger. While the $\ell=2, m=2$ GW mode is the primary contributor to the
GW signal, it is damped faster than the other modes leading to
interesting secular behaviours.

\begin{figure*}
  \includegraphics[width=0.48\textwidth]{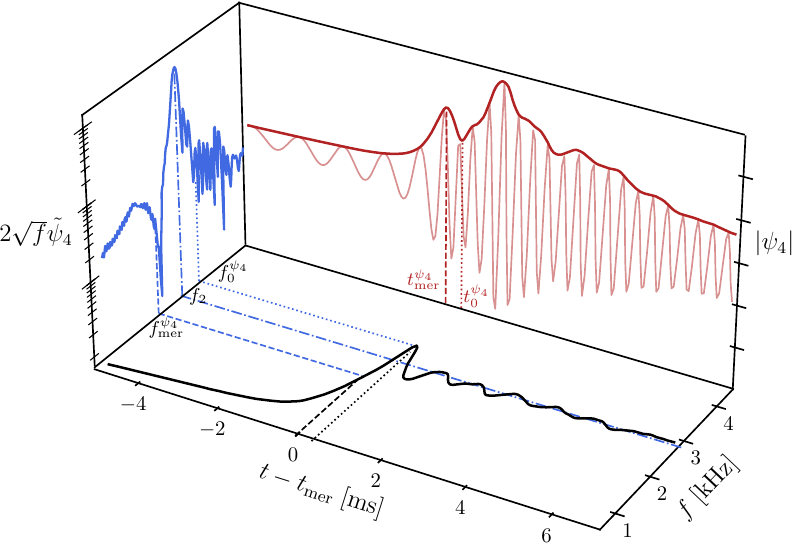}
  \hskip 0.5cm 
  \includegraphics[width=0.48\textwidth]{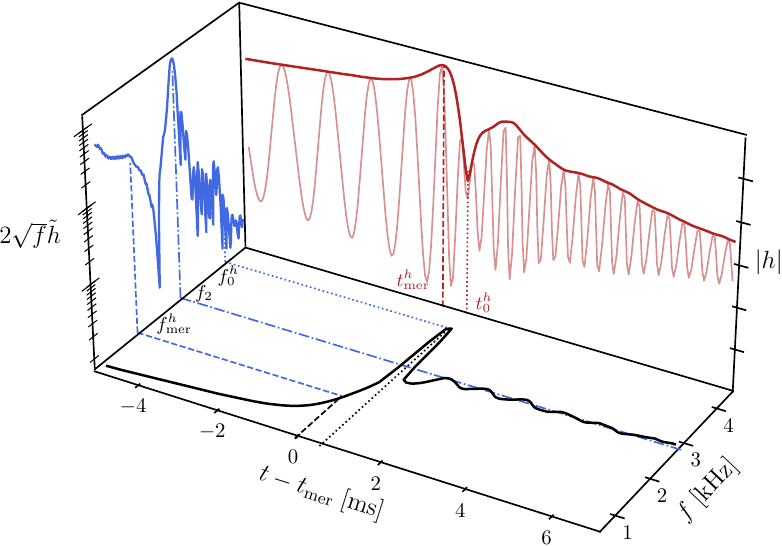}
  \caption{3D representation of the complete information in the GW signal
    from a representative binary~\citep{Tootle2022}. The left panel shows
    the $\ell=2,m=2$ mode of the GW signal $\psi_{4}(t)$ (light red) and
    its amplitude $|\psi_{4}(t)|$ (dark red), the instantaneous frequency
    $f_{_{\rm GW}}(t)$ (black), and the power spectral density (PSD)
    $\sqrt{2}f\, \tilde{\psi}_{4}$ (blue) as a function of the frequency
    $f$. Also indicated are the frequency at merger $f_{\rm
      mer}^{\psi_{4}}$, the frequency at quasi time-symmetry
    $f^{\psi_{4}}_{0}$, the dominant frequency of the HMNS emission
    $f_{2}$, and the corresponding times where these frequencies
    appear. The right panel shows the same quantities but when computed
    in terms of the GW strain.}
   \label{fig:3D_PSD_amp_freq}
\end{figure*}

We here take a new look at both of these old problems by considering the
spectral properties of the GW signal when computed in terms of the Weyl
scalar $\psi_4$. In this way, we are able to find three novel features
that can enrich our understanding of the GW signal from BNS mergers. In
particular, we first highlight the presence of a new {instantaneous
  frequency}, which we dub $f^{\psi_4}_0$, that can be associated with
the instant of quasi time-symmetry in the postmerger
dynamics. Interestingly, we find that a quasi-universal relation exists
for $f^{\psi_4}_0$ as a function of the tidal deformability
$\kappa_{2}^{_{T}}$ and of the binary mass ratio $q$. Second, by
employing a large number of BNS simulations, some of which are taken from
the \texttt{CoRe} database~\citep{Gonzalez2022}, we obtain a new
quasi-universal relation for $f_{\rm mer}$ as a function of
$\kappa_{2}^{_{T}}$ and $q$ that not only requires a smaller number of
coefficients, but also provides a more accurate description of the
data. Finally, as already suggested in~\citep{Papenfort:2022ywx}, we
provide evidence that the $\ell=2, m=1$ GW mode could become the most
powerful mode on secular timescales after the merger.

\section{Numerical and physical framework}
\label{sec:num_phys_framework}

Our analysis is based on the GW signal computed via numerical simulations
of BNS mergers in full general relativity computed with the codes
described in~\citep{Radice2013b, Radice2013c, Most2019b, Most2019,
  Papenfort2021b, Tootle2021} and using a number of different EOSs (see
below). In addition, we employ part of the data contained in the
\texttt{CoRe} database \citep{Gonzalez2022}, from where we select only
simulations with the highest-resolution. The combined data of $118$
irrotational binaries covers the range $q:=M_2/M_1\in [0.485,1]$ in the
mass ratio, $M:=M_{1}+M_{2} \in [2.4,3.33]~M_{\odot}$ in the total ADM
mass at infinite separation, and $\kappa_{2}^{_{T}} \in [33, 458]$ in the
tidal deformability. The dataset comprises a variety of EOSs including
some with quark matter~\citep{Prakash:2021wpz, Logoteta2021, Alford2005,
  Demircik:2021zll, Tootle2022}.

A crucial role in our analysis is played by the use of the Weyl scalar
$\psi_4$ in place of the standard dimensionless strain polarisations
$h_{+,\times}$. The two quantities are mathematically equivalent and
related by two time derivatives (\ie $\psi_4 = \partial^2_t (h_+ -
ih_{\times})$; see~\citep{Bishop2016} for a review). However, while
$\psi_4$ is computed from the simulations, $h_{+,\times}$ are obtained
after a nontrivial double time integration (the transformation from
$h_{+,\times}$ to is $\psi_4$ trivial as it involves derivatives and not
integrals; see~\citep{Bustillo2022} for a data-analysis framework based
on $\psi_4$, which can obviously be employed for all types of
compact-object binaries). More importantly, the evolution of the GW
frequency from $\psi_4$ is less rapid than from the strain, \ie
$\partial_t \ln f^{\psi_4}_{_{\rm GW}}(t) \ll \partial_t \ln f^{h}_{_{\rm
    GW}}(t)$, thus making it easier and more robust to characterise the
features of the $\psi_4$ GW signal. In this sense, while $\psi_4$ and
$h_{+,\times}$ are related by simple time derivatives, the analysis
carried out with the former does provide additional information as it
allows for the determination of {properties} that are harder to
capture with the latter.

\section{Old and new {frequencies}}
\label{sec:old_and_new_spectral}

Figure~\ref{fig:3D_PSD_amp_freq} reports the complete information of the
GW signal from a representative binary in our sample. Using a 3D
representation, we report on the left the $\ell=2,m=2$ mode of the GW
signal $\psi_{4}(t)$ (light red) and its amplitude $|\psi_{4}(t)|$ (dark
red), the instantaneous frequency $f^{\psi_4}_{_{\rm GW}}(t)$ (black),
and the power spectral density (PSD) $\sqrt{2}f\, \tilde{\psi}_{4}$
(blue) as a function of the frequency $f$~\citep[see][for details on the
  definition]{Rezzolla2016}. Also indicated are the three main
frequencies in our analysis: the frequency at merger $f_{\rm
  mer}^{\psi_{4}}$, \ie the GW frequency at the \textit{first} maximum of
$|\psi_{4}|$, the frequency at quasi time-symmetry $f^{\psi_{4}}_{0}$,
\ie the GW frequency at the \textit{first} minimum of $|\psi_{4}|$, and
the dominant frequency of the HMNS emission $f_{2}^{\psi_{4}}$. To help
the eye, we also mark with lines the corresponding times $t^{\psi_4}_{\rm
  mer}$ (dashed), $t^{\psi_4}_{0}$ (dotted), and frequencies (dashed,
dotted and dot-dashed respectively). The right panel of
Fig.~\ref{fig:3D_PSD_amp_freq} shows the same quantities but when
computed from the strain. {By comparing the black lines in the left and
  right panels it is straightforward to realise that the variation of
  $f^{h}_{_{\rm GW}}(t)$ is much larger than that in $f^{\psi_4}_{_{\rm
      GW}}(t)$ over the {same interval of $\sim 1\,$ms after the
    merger}. It is this very rapid change in $f^{h}_{_{\rm GW}}(t)$ that
  makes the identification of {$f^{h}_{0}$} extremely difficult, if not
  impossible. Note also that} while in both representations $f_{\rm mer}
< f_{2} < f_{0}$, the numerical values of the various quantities are
similar but not identical. However, $f_{2}^{\psi_{4}} \simeq f_{2}^{h}$
to very good precision (the largest differences are $\lesssim4\%$) simply
because this frequency is relative to a mostly monochromatic GW signal;
hence, hereafter we simply assume $f_{2}^{\psi_{4}} = f_{2}^{h} =:
f_2$. {Finally, because in all representations $f_0$ is largest frequency
  measured, even a crude measure of largest frequency in the signal will
  serve as a first estimate of the $f_0$ frequency.}\footnote{From a
  numerical point of view, we note that the $f_0$ frequency is always
  below $\sim 4\,{\rm kHz}$ and is therefore much smaller than the
  typical sampling frequency of the $\psi_4$ scalar, that is $\simeq
  80\!-\!100\,{\rm kHz}$.}

Besides marking the time of the first amplitude minimum, from a physical
point of view $t^{\psi_4}_{0}$ corresponds to the time when the two
stellar cores have reached the minimum separation and are about to
bounce-off each other. At this instant, the corresponding amplitude of
$\psi_4$ shows a clear minimum, while the instantaneous GW frequency a
local maximum [the discussion in the \textit{Supplemental Material} (SM)
  illustrates this behaviour very clearly by employing the toy model
  introduced in~\citep{Takami2015}].

\begin{figure*}
  \centering 
  \includegraphics[height=0.04\textwidth]{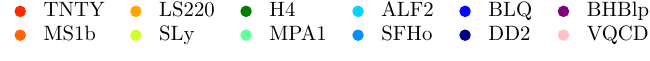}\\  
  \includegraphics[width = 0.48\textwidth]{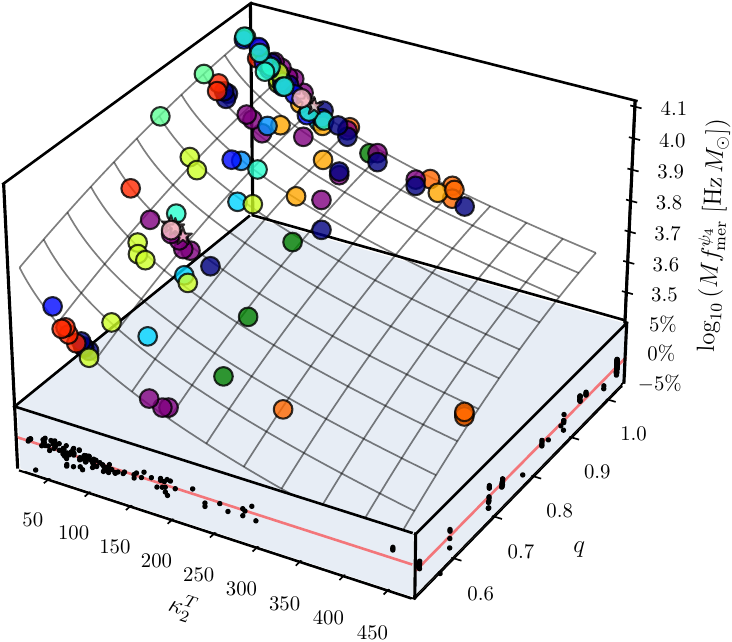}
  \hskip 0.2cm
  \includegraphics[width=0.48\textwidth]{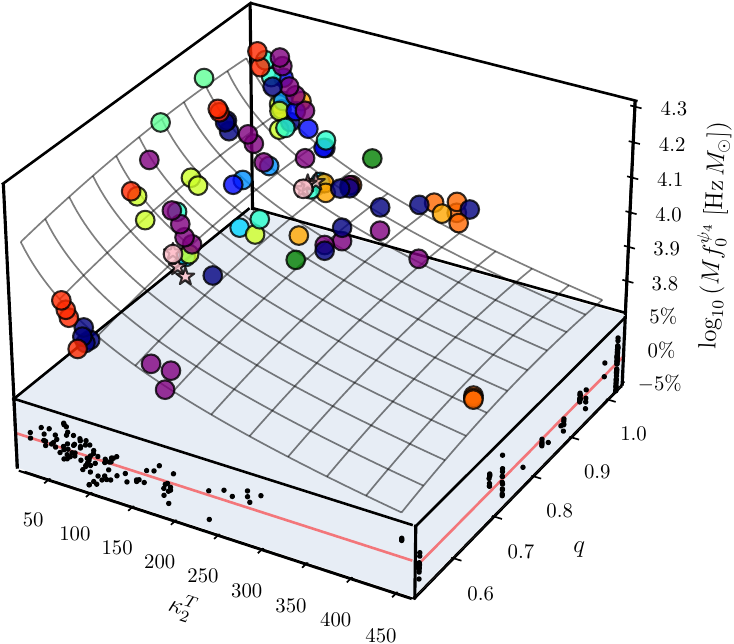}
  \caption{3D representation of the GW frequencies $f^{\psi_{4}}_{\rm
      mer}$ and $f^{\psi_{4}}_{0}$ as measured from the data (coloured
    circles) and presented as a function of $\kappa^{_{T}}_2$ and
    $q$. Also reported are the best-fit surfaces, while shown below are
    the relative errors of the fit in the two principal directions. Stars
    mark binaries modelled with the V-QCD EOS and thus having a strong
    first-order phase transition~\citep{Demircik:2021zll, Tootle2022}.}
    \label{fig:2}
\end{figure*}

\section{Quasi-universal relations}
\label{sec:quasi-universal_relations}

We next proceed to the derivation of quasi-universal relations that can
be employed to deduce the physical properties of the binary. Following
the approach started already in~\citep{Takami2014, Takami2015,
  Rezzolla2016}, which captures the logarithmic variation of a properly
rescaled mass and frequency, we express the relevant frequencies in terms
of a power expansion of the mass ratio $q$, \ie
$\log_{10}\left[({M}/M_{\odot}) ({f}/\rm{Hz}) \right] = a_{0} + (b_{0} +
b_{1}q + b_{2}q^{2}) \left({\kappa_{2}^{_{T}}}\right)^n$, where $f$ is
any of the frequencies we consider (\ie $f^{\psi_{4}}_{\rm mer},
f^{h}_{\rm mer}, f^{\psi_{4}}_{0}, f_{2}$), $a_{0}, b_{0}, b_{1}, b_{2},
n$ are fitting coefficients. Hereafter, we will refer to this generic
fitting functions as $\mathcal{F}_{1}$.

Figure~\ref{fig:2} provides a 3D representation of the measured GW
frequencies $f^{\psi_{4}}_{\rm mer}$ and $f^{\psi_{4}}_{0}$ as a function
of $\kappa^{_{T}}_2$ and $q$ (see also the SM for fits to $f^{h}_{\rm
  mer}$ and $f_2$). Also reported is the fitting surface described by
$\mathcal{F}_1$, with the best-fit parameters listed in Table~$2$ of the
SM for all the frequencies considered. Furthermore, for each frequency we
report below the relative error of the fit in the two principal
directions of the fit, $\kappa^{_{T}}_2$ and $q$. Despite their simple
form, our fits for $f^{\psi_{4}}_{\rm mer}$ and $f^{\psi_{4}}_{0}$
capture the data very well, showing average relative errors that are
$\lesssim 1\%$ and maximal relative errors $\lesssim 2\%$ for other than
equal-mass binaries.

\begin{figure}
\includegraphics[width=0.98\columnwidth]{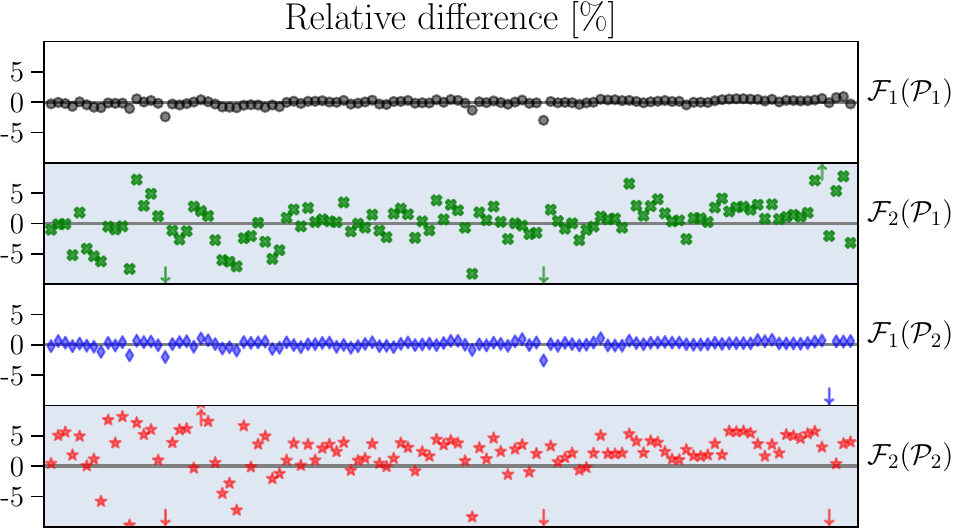}
\caption{Relative differences between the measured $f_{\rm mer}^{h}$
  frequencies and the corresponding values from the fit. Different rows
  refer to the four different possibilities of applying the fitting
  function $\mathcal{F}_{i}$ to the data pipeline $\mathcal{P}_{i}$ with
  $i = 1,2$; arrows indicate relative differences above $10\%$. Note how
  the new fitting function and pipeline, \ie
  $\mathcal{F}_{1}(\mathcal{P}_{1})$, provide errors that are about four
  times smaller.}
\label{fig:fmax_core_vs_us}
\end{figure}

It is interesting to compare our functional fitting form $\mathcal{F}_1$
for $f^{h}_{\rm mer}$, which needs only five fitting coefficients, with
the one proposed in~\citep{Breschi2022} for irrotational binaries, which
we will refer to as $\mathcal{F}_{2}$, and that requires twice as many
coefficients. In order to compare $\mathcal{F}_{1}$ and $\mathcal{F}_{2}$
it is first necessary to distinguish the ``pipeline'', that is, the
technical procedure employed to extract the frequencies from the data. We
thus indicate with $\mathcal{P}_{1}$ the pipeline discussed above and
with $\mathcal{P}_{2}$ that released in~\citep{Gonzalez2022}. Naturally,
each fitting function can be applied to either pipeline, so that
$\mathcal{F}_1(\mathcal{P}_1)$ indicates the use of our fitting form to
data computed with our pipeline. In Fig.~\ref{fig:fmax_core_vs_us}, we
present the relative differences between the measured frequencies for the
118 binaries considered and the corresponding values from the fit, with
different rows referring to the four possibilities.

Overall, the comparison in Fig.~\ref{fig:fmax_core_vs_us} shows that
$\mathcal{F}_{1}$ leads to smaller relative errors with a maximum
residual error of $\sim2\%$ and an an average residual error that is
between two and four times smaller than for $\mathcal{F}_2$. As a
cautionary note we should remark that we have specialised the fitting
$\mathcal{F}_2$, which is more general and can include spinning and
eccentric binaries, to the case relevant for this comparison, namely,
irrotational binaries. Hence, our conclusions apply only to such
binaries.

\begin{figure*}
\includegraphics[width=0.48\textwidth]{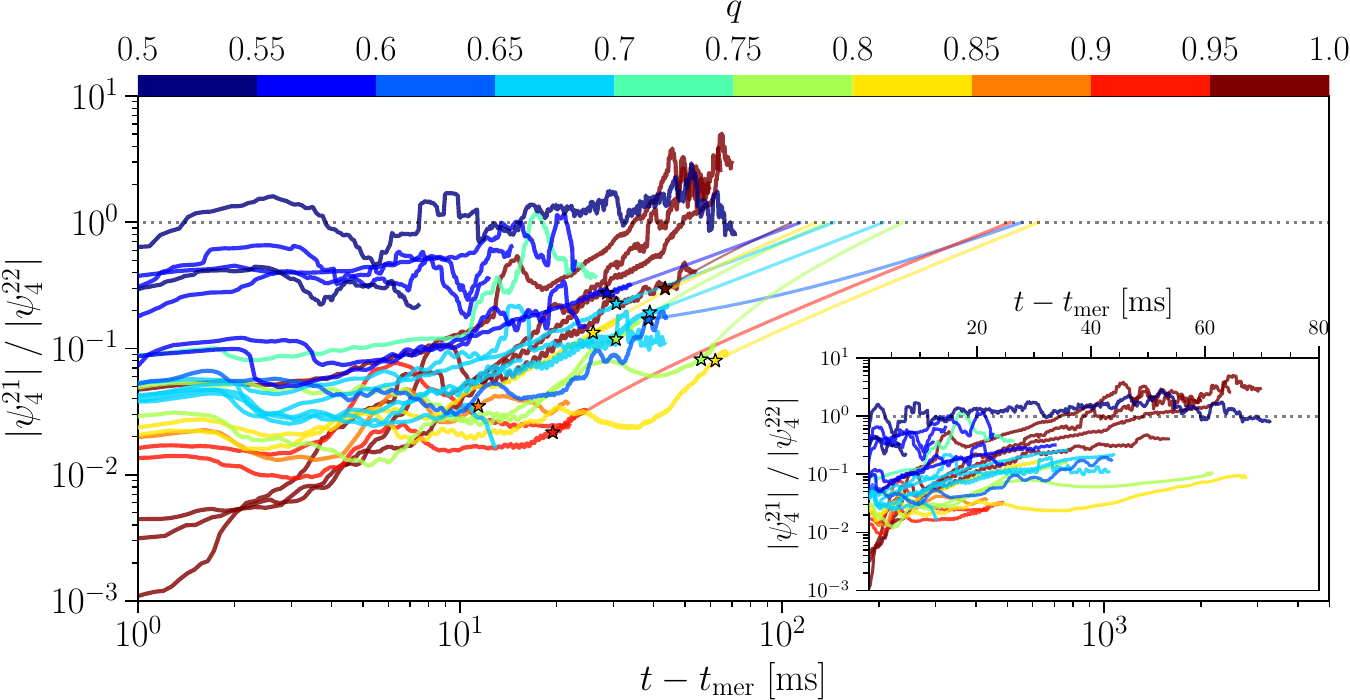}
\includegraphics[width=0.48\textwidth]{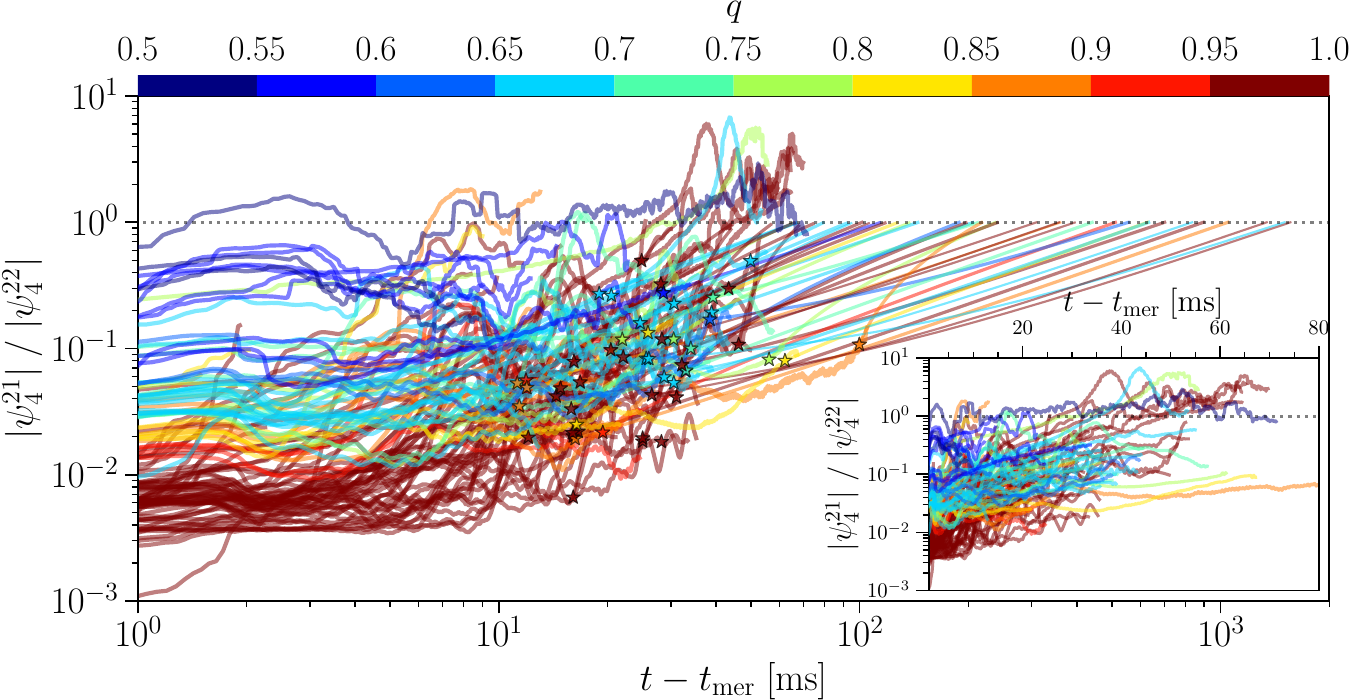}
\caption{Evolution of the ratio of the two main GW modes
  $|\psi_{4}^{21}|/|\psi_{4}^{22}|$, with the left panel showing a
  selected set of binaries to highlight the behaviour for different mass
  ratios (colour code) and the right one reporting all of the
  binaries. Black dotted horizontal lines are used to mark the position
  when the two modes have equal amplitudes, while straight coloured lines
  report a linear extrapolation after averaging the last $5\, {\rm ms}$
  of the evolution; note that the large majority of binaries exhibits a
  growing trend that is maintained throughout the computed evolution.}
\label{fig:psi4amp_ratio}
\end{figure*}

\section{Secular GW emission}
\label{sec:secular_gw_emission}

The last point we cover regards the relative strengths of the $\ell=2,
m=2$ and $\ell=2, m=1$ GW modes. The importance of the latter was first
pointed out in~\citep{Paschalidis2015, Lehner2016} and is produced by
corresponding asymmetries in the rest-mass density. The emergence of an
$m=1$ deformation is well-known to occur in isolated
stars~{\citep{Chandrasekhar1969book, Shibata:2000jt, Baiotti06b,
    Franci2013b, Loeffler2015}} that have a sufficiently large amount of
rotational kinetic energy $T$ and emerges when the ratio $T/|W|$, where
$W$ is the gravitational binding energy, exceeds a certain threshold
{\citep[in a systematic analysis, ][have shown that this happens for
    $T/|W| \gtrsim0.25$]{Baiotti06b}}. In such stars, the $m=1$ mode in
the rest-mass density would grow exponentially reaching equipartition
with the $m=2$ mode, and, subsequently, represent the largest
deformation. A similar phenomenology seems to be present also for the
postmerger remnant, as already hinted in~\citep{Papenfort:2022ywx}, but
as shown more clearly by the numerous binaries considered
here. Figure~\ref{fig:psi4amp_ratio} reports the evolution of the ratio
of the GW amplitudes in the two modes $|\psi_{4}^{21}|/|\psi_{4}^{22}|$,
with the left panel showing a selected set of binaries and with the right
panel reporting all binaries (the raw timeseries are smoothed over a
window $\Delta t=0.5\,{\rm ms}$). While only a few binaries in the sample
reach $|\psi_{4}^{21}|/|\psi_{4}^{22}|=1$ within the simulated time, the
large majority exhibits a trend that we try to capture by extrapolating
linearly in time after averaging the last $5\, {\rm ms}$ of the
evolution. Using a colour code to distinguish binaries with different
$q$, it becomes clear that the initial strength of the $m=1$ mode is
inversely proportional to the mass ratio, so that for an extremely
asymmetric binary, \ie $q \lesssim 0.6$,
$|\psi_{4}^{21}|/|\psi_{4}^{22}|$ can be more than two orders of
magnitude larger than for equal-mass binaries. At the same time, the
initial mode-amplitude ratio does not depend on the merger dimensionless
spin~\citep{Papenfort:2022ywx}. {Unsurprisingly, a similar behaviour
  can be observed when computing the signal-to-noise (SNR) ratio in the
  $m=2$ and $m=1$ modes. Specifically, by computing a time-windowed SNR
  ratio we can quantify the growing contribution of the subdominant mode
  in a similar fashion to estimates given by \citep{Lehner2016a} and find
  that its increases to $\simeq \mathcal{O}(1)$ if the $m=1$ is not
  suppressed (see SM for details).}

If confirmed by systematic, long-term evolutions, this finding would
change the standard picture in which the largest signal in the BNS
postmerger is to be expected at the $f_2$ frequency. Rather, the trend
reported here suggests that the most powerful feature in the PSD for
long-lived HMNSs may actually appear at a frequency $\simeq
\tfrac{1}{2}f_2$. Because this falls in a more favourable region of the
detectors sensitivities, and assuming a detection angle not favouring
either of the modes, the corresponding signal-to-noise ratio will grow
proportionally to the ratio between detectors noise at $\tfrac{1}{2} f_2$
and $f_2$, hence by a factor of $\sim 2$ for LIGO or Virgo.

While promising, this prospect should be accompanied by some
caveats. First, it is possible that the growth rate may be weaker than
the one estimated here. Second, the extrapolation assumes that the HMNS
will not collapse to a black hole before reaching
$|\psi_{4}^{21}|/|\psi_{4}^{22}| \sim 1$ and while this is likely for
soft EOSs and low-mass binaries, it may not happen if the EOS is stiff
and the binary massive. Third, all binaries in our sample have zero
deformation. A robust conclusion that can be inferred from the
  results shown in Fig.~\ref{fig:psi4amp_ratio} is that remnants with a
  long lifetime, as it was likely the case for
  GW170817~\citep{Rezzolla2017, Gill2019, Murguia-Berthier2021}, will
  reasonably have the $\ell=2, m=1$ as the least-damped mode. Hence,
  considerable spectral power should be present at frequencies
  $\frac{1}{2}f_2$ and $f_2$, with the main strain amplitudes in a ratio
  $|h^{21}|/|h^{22}| \sim 0.1-1$ for generic orientations (\eg for an
  inclination of $2\arctan(1/2)\sim 53^{\circ}$ two modes have the same
  spin-weighted spherical-harmonics coefficients).

\section{Conclusion}
\label{sec:conclusion}

{Leveraging on a rich literature developed over the last ten years on
  this subject,} we have considered again the spectral properties of the
signal when computed in terms of the Weyl scalar $\psi_4$ rather than in
terms of the GW strain $h_{+,\times}$. Exploiting the better behaviour of
$\psi_4$, we were able to highlight three novel features that can be used
to better infer physical information from the detected signal.

First, by employing a large number of simulations spanning a considerable
set of EOSs and mass ratios, we have shown the existence of a new
{instantaneous frequency}, $f^{\psi_4}_0$, that can be associated
with the instant of quasi time-symmetry in the postmerger dynamics. This
corresponds to when the stellar cores in the merger remnant have reached
their minimum separation and are about to bounce-off each other. Just
like other spectral frequencies of the BNS GW signal, $f^{\psi_4}_0$ also
follows a quasi-universal behaviour as a function of the tidal
deformability $\kappa_{2}^{_{T}}$ and of the binary mass ratio $q$, for
which we provide a simple and yet accurate analytical expression. Second,
we have obtained a new quasi-universal relation for the merger frequency
$f^{h}_{\rm mer}$ as a function of $\kappa_{2}^{_{T}}$ and $q$. The new
expression not only requires a smaller number of fitting coefficients
than alternative expressions in the literature, but it also provides a
more accurate description of the data, with a residual error that is four
times smaller on average. Finally, we have pointed out the evidence that
the $\ell=2, m=1$ could become the most powerful GW mode on sufficiently
long timescales, with strain amplitudes for the dominant modes that are
in a ratio $|h^{21}|/|h^{22}| \sim 0.1-1$. Should this mode not be
suppressed by the collapse of the HMNS to a black hole or by other
dissipative effects such as magnetic fields, considerable spectral power
should be present at frequencies $\frac{1}{2}f_2$, where it could be
detected in conditions of smaller signal-to-noise ratios or by
third-generation detectors.

The results presented here can be improved by enlarging the number of BNS
simulations considered, by increasing the variance in the microphysical
description (\eg including simulations with magnetic fields and neutrino
transport), by performing additional long-term evolutions, and by
extending the fitting approach to binaries with spins and
eccentricity. We will explore these extensions in future work.

\noindent
\textit{Data policy.} The relevant data that supports the findings of
this paper is available from the first author and can be shared upon a
reasonable request.

\smallskip
\noindent
We thank K. Chakravarti, K. Takami, C. Ecker, and C. Musolino for useful
input and discussions. Support comes from the State of Hesse within the
Research Cluster ELEMENTS (Project ID 500/10.006). LR acknowledges
funding by the ERC Advanced Grant ``JETSET: Launching, propagation and
emission of relativistic jets from binary mergers and across mass
scales'' (Grant No. 884631). The simulations from which parts of the used
data are derived were performed on HPE Apollo HAWK at the High
Performance Computing Center Stuttgart (HLRS) under the grants BNSMIC and
BBHDISKS, and on SuperMUC at the Leibniz Supercomputing Centre.

\software{\texttt{Einstein Toolkit} \citep{EinsteinToolkit_etal:2020_11},
	\texttt{Carpet} \citep{Schnetter:2003rb},
	\texttt{FIL} \citep{Most2019b},
	\texttt{FUKA} \citep{Papenfort2021b},
	\texttt{Kadath} \citep{Grandclement09},
	\texttt{Watpy} \citep{Watpy},
	\texttt{Kuibit} \citep{kuibit21},
	\texttt{Mathematica} \citep{Mathematica}
}

\bibliographystyle{aasjournal}

\appendix 
\section*{Supplemental Material}

\subsection*{Toy-model analogy}

Although it is possible to characterise the new spectral feature $f_{0}$
simply in terms of the instantaneous GW frequency at the time
$t^{\psi_4}_{0}$ at which the Weyl scalar has its first minimum, it is
more interesting to associate with this definition also a physical
interpretation. As mentioned in the main text, $t^{\psi_4}_{0}$
effectively corresponds to when the two stellar cores have reached the
minimum separation and are about to bounce-off each other. At this
instant, the radial velocity of the two stellar cores and the angular
velocity of the HMNS has a minimum, so that the corresponding amplitude
of $\psi_4$ has a minimum and instantaneous GW frequency a local maximum.
 
To illustrate this behaviour, we employ the toy model developed
in~\cite{Takami2015} (and also employed in other studies,
\eg~\cite{Ellis2018b, Lucca2021}) to describe the dynamics of the
postmerger remnant and which consists of an axisymmetric disk rotating
rapidly at a given angular frequency, say $\Omega_2$, to which two
spheres reproducing the two stellar cores are connected but are also free
to oscillate via a spring that connects them (see
Fig.~17~\cite{Takami2015}). In such a system, the two spheres will either
approach each other, decreasing the moment of inertia of the system, or
move away from each other, increasing the moment of inertia. Because the
total angular momentum is essentially conserved, the system's angular
frequency will vary between a minimum value $\Omega_1$ (corresponding to
the time when the two spheres are at the largest separation) and a
maximum value $\Omega_3$ (corresponding to the time when the two spheres
are at the smallest separation). Overall, the mechanical toy model will
rotate with an angular frequency that is a function of time and bounded
by $\Omega_1$ and $\Omega_3$, where more time is spent and hence more
spectral power is accumulated. If a damping is introduced, the excursion
of the oscillations between the spheres will decrease over time and
eventually stop; when this happens, the toy model will simply rotate at
the frequency $\Omega_2$, as does the HMNS after the transient period and
before other deformations (\eg the $\ell=2,m=1$ mode) affect it.
 
The Lagrangian of the toy model can be found in~\cite{Takami2014} and
from it it is possible to derive the differential equation for the radial
displacement $r(t)$
 \begin{equation}
   \ddot{r} + \frac{4k(r-r_{0})}{m} - \Bigg{[} \frac{c_{1}}{r^{2} +
       M R^{2}/(2m)} \Bigg{]}^{2} r + \frac{2br}{m} = 0\,,
   \label{eq:app:toy_model_ode}
 \end{equation}
where $M$ and $m$ are the masses of the disc and of the spheres, $R$ is
the radius of the disc, $c_{1}$ an integration constant related to the
total angular momentum, $r_{0}$ the natural displacement of mass and $b$
is responsible for dissipative effects. The dissipation due to the
emission of GWs and which results in the two spheres getting closer to
one another with time and with a decreasing radial displacement is
introduced by varying the natural displacement $r_{0}$ via via an
exponential function $\exp({-{t}/{\tau}})$ with a suitably chosen
relaxation time $\tau$. Note that this modification carries straight from
the Lagrangian to the final ODE without additional time derivatives, as
it is present in the derivative of the Lagrangian with respect to the
canonical coordinate $r(t)$ and not its conjugate.
 
The system is sufficiently simple that it is possible to compute the GW
amplitude and frequency derived from the quadrupole formula so that the
strain components read
 \begin{equation}
   h_{+} = \frac{2m}{d} [\{ \dot{r}^{2} + r(\ddot{r}-2r\Omega^{2}) \}\cos(2\varphi) 
     - r ( 4\dot{r}\Omega + r\dot{\Omega} )\sin(2\varphi)~]\,, \nonumber
   \label{eq:app:hpp_toy_model}
 \end{equation}
 \begin{equation}
     h_{\times} = \frac{2m}{d} [ \{ \dot{r}^{2} + r(\ddot{r}-2r\Omega^{2}) \}\sin(2\varphi) 
      + r ( 4\dot{r}\Omega + r\dot{\Omega} )\cos(2\varphi)~]\,, \nonumber
         \label{eq:app:hxx_toy_model}
 \end{equation}
with $d$ the distance from the source to the detector. The $\psi_{4}$
polarisations are obtained by differentiating twice with respect to
time. Introducing the abbreviations $A\,\coloneqq \dot{r}^{2} +
r(\ddot{r}-2r\Omega^{2})$ and $B\, \coloneqq r ( 4\dot{r}\Omega +
r\dot{\Omega} ) $, as well as recalling that $\Omega(t) =
\dot{\varphi}(t)$ we write them now explicitly
 \begin{equation}
   \psi_{4,+} = \frac{2m}{d} [  -\sin(2\varphi)\{ 4\dot{\varphi}(\dot{A} - B\dot{\varphi})
     + \ddot{B} + 2A\ddot{\varphi}  \} 
      + \cos(2\varphi)\{ -4\dot{\varphi}(\dot{B}+A\dot{\varphi} )
     + \ddot{A} - 2B\ddot{\varphi} \}  ~]\,, \nonumber
   \label{eq:app:psi4pp_toy_model}
 \end{equation}
 \begin{equation}
   \psi_{4,\times} = -\frac{2m}{d} [  \cos(2\varphi)\{ 4\dot{\varphi}(\dot{A} - B\dot{\varphi})
     + \ddot{B} + 2A\ddot{\varphi}  \}  
      + \sin(2\varphi)\{ -4\dot{\varphi}(\dot{B}+A\dot{\varphi} )
     + \ddot{A} - 2B\ddot{\varphi} \}  ~]\,, \nonumber
   \label{eq:app:psi4xx_toy_model}
 \end{equation}
where $\psi_{4,+} := \partial^2_t h_{+}$ and $\psi_{4,\times} :=
\partial^2_t h_{\times}$.
 
The amplitude and the instantaneous frequency of the signal are
calculated using the usual definitions
 \begin{eqnarray}
    \vert \psi_{4}(t) \vert &\coloneqq& \sqrt{ \psi_{4,\times}^{2}(t) + \psi_{4,+}^{2}(t)}\,,\\
    f_{\rm GW}^{\psi_{4}}(t) &\coloneqq& \frac{1}{2 \pi} \frac{d}{dt}\arctan 
    \left(\frac{\psi_{4,\times}(t)}{\psi_{4,+}(t)}\right)\,.
 \end{eqnarray}
 
\begin{figure*}
     \includegraphics[width=0.48\textwidth]{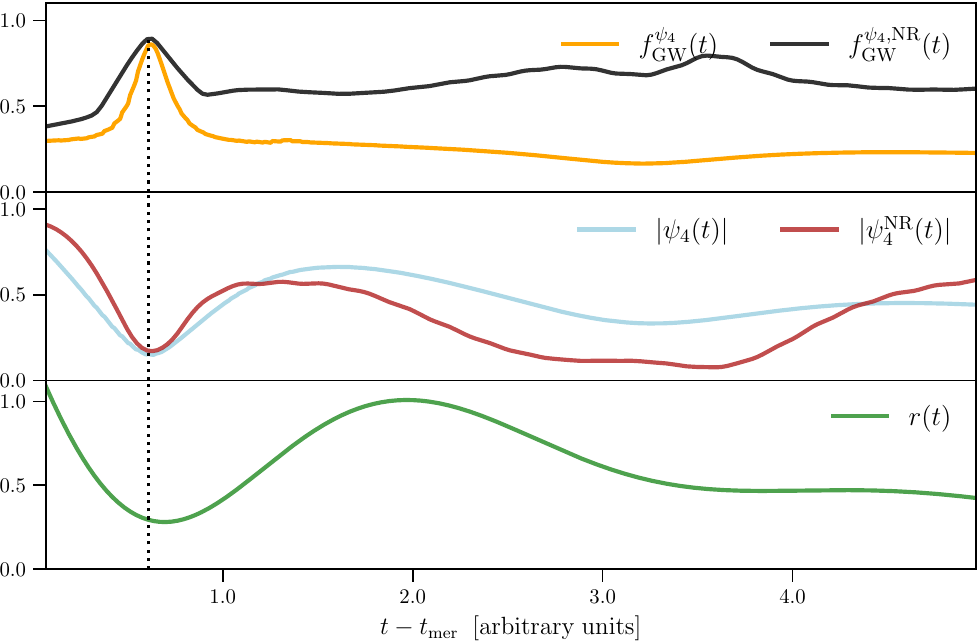}
         \hskip 0.5cm
         \includegraphics[width=0.48\textwidth]{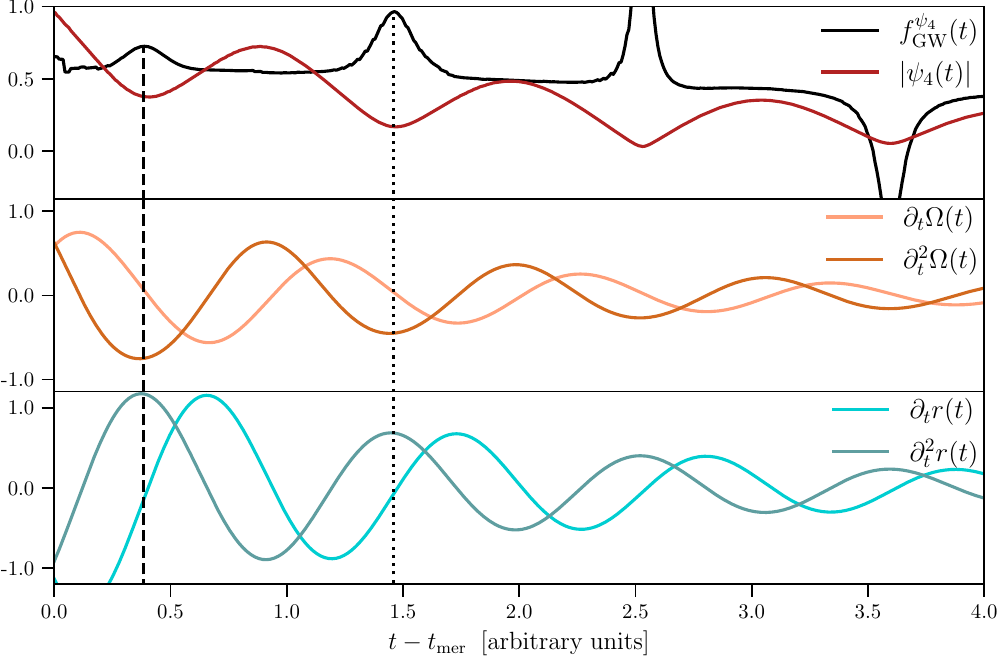}
         \caption{Solution of the toy model for a representative set of
           parameters. The left panel reports the GW frequency (top row)
           from the toy model and from an actual simulation (marked by the
           index ``NR''), the GW amplitudes (middle row) and the radial
           displacement $r(t)$ (bottom row). The right panel shows instead
           the relation between the GW quantities and other dynamical
           quantities of the toy model, such as the time derivatives of
           the angular velocity and of the radial displacement. Note that
           the GW frequencies and amplitudes are anti-correlated and the
           minimum in the radial displacement at $t^{\psi_{4}}_{0}$
           corresponds to the maximum frequency $f^{\psi_{4}}_{0}$ and
           minimum amplitude. Also, the minima of the amplitude (maxima of
           the frequency) correspond to moments in time when the angular
           frequency $\Omega$ attains a local maximum, and the radial
           displacement a local minimum.}
    \label{fig:toy_model}
 \end{figure*}
 
The rather lengthy final expressions read:
 %
 \begin{eqnarray}
 \frac{d}{2m} \vert \psi_{4}(t) \vert &=& \big{[}~4
   \sin(2\varphi)\dot{A}\dot{\varphi} +
   4\cos(2\varphi)\dot{B}\dot{\varphi} + 4A\cos(2\varphi)\dot{\varphi}^{2}
   - 4B\sin(2\varphi)\dot{\varphi}^{2} - \cos(2\varphi)\ddot{A} +
   \sin(2\varphi)\ddot{B} \\  \phantom{=}&& +2 B\cos(2\varphi)\ddot{\varphi} +
   2A\sin(2\varphi)\ddot{\varphi} ~\big{]}^{2} + \big{[}~
   4\cos(2\varphi)\dot{A}\dot{\varphi} -
   4\sin(2\varphi)\dot{B}\dot{\varphi} - 4B\cos(2\varphi)\dot{\varphi}^{2}
   -4A \sin(2\varphi)\dot{\varphi}^{2} \nonumber \\ \phantom{=}&&+
   \sin(2\varphi)\ddot{A} + \cos(2\varphi)\ddot{B} +
   2A\cos(2\varphi)\ddot{\varphi} - 2B\sin(2\varphi)\ddot{\varphi}
   ~\big{]}^{2}, \nonumber \\ 2\pi \frac{f_{\rm GW}^{\psi_{4}}}{F^{\psi_{4}}} &=&
 48\dot{A}^{2}\dot{\varphi}^{3} + 48\dot{B}^{2}\dot{\varphi}^{3} + 32
 A^{2} \dot{\varphi}^{5} + 32 B^{2} \dot{\varphi}^{5} -
 32A\dot{\varphi}^{3}\ddot{A} + 6\dot{\varphi}\ddot{A}^{2} - 32
 B\dot{\varphi}^{3}\ddot{B} + 6\dot{\varphi}\ddot{B}^{2} - 24 B
 \dot{\varphi}\ddot{A}\ddot{\varphi} \\ \phantom{=}&&+ 24A\dot{\varphi}\ddot{B}\ddot{\varphi}
 + 24 A^{2} \dot{\varphi} \ddot{\varphi}^{2} + 24
 B^{2}\dot{\varphi}\ddot{\varphi}^{2} + 4 B\dot{\varphi}^{2}\dddot{A} -
 \ddot{B}\dddot{A} - 2A\ddot{\varphi}\dddot{A} -
 4A\dot{\varphi}^{2}\dddot{B} + \ddot{A}\dddot{B} -2 B \ddot{\varphi}
 \dddot{B} \nonumber \\ \phantom{=}&&+ 2\big{(}-4A^{2}\dot{\varphi}^{2} + A\ddot{A} +
 B(-4B\dot{\varphi}^{2} +\ddot{B}) \big{)}\dddot{\varphi} +
 2\dot{B}\big{(}3\ddot{B}\ddot{\varphi} +
 6\dot{\varphi}^{2}(-3\ddot{A}+4B\ddot{\varphi}) - 2\dot{\varphi}\dddot{B}
 + A(40\ddddot{\varphi} + 6\ddot{\varphi}^{2} -
 4\dot{\varphi}\dddot{\varphi}) \big{)} \nonumber \\ \phantom{=}&&+ \dot{A}\big{(} 6
 \ddot{A}\ddot{\varphi} + 12\dot{\varphi}^{2}(3\ddot{B}+4A\ddot{\varphi})
 - 4\dot{\varphi}\dddot{A} -4B(20 \dot{\varphi}^{4} + 3\ddot{\varphi}^{2}
 -2\dot{\varphi}\dddot{\varphi}) \big{)}, \nonumber  \\ \frac{-1}{F^{\psi_{4}}} &:=& 16
 \dot{A}^{2}\dot{\varphi}^{2} + 16 \dot{B}^{2}\dot{\varphi}^{2} +
 16A^{2}\dot{\varphi}^{4} + 16B^{2}\dot{\varphi}^{4}
 -8A\dot{\varphi}^{2}\ddot{A} + \ddot{A}^{2} - 8B\dot{\varphi}^{2}\ddot{B}
 + \ddot{B}^{2} + 4(-B\ddot{A} + A\ddot{B})\ddot{\varphi} \\ \phantom{=}&&+
 4(A^{2}+B^{2})\ddot{\varphi}^{2} +
 8\dot{A}\dot{\varphi}(-4B\dot{\varphi}^{2} + \ddot{B} + 2A\ddot{\varphi})
 +8\dot{B}\dot{\varphi}(4A\dot{\varphi}^{2} - \ddot{A} +
 2B\ddot{\varphi})\,. \nonumber
 \end{eqnarray}

Figure~\ref{fig:toy_model} reports the solution of the system when
considering the following set of representative parameters: $M=m=10$,
$k=0.2$, $b=0.25$, $c_{1}=25$, $R=22$, $r_{0}=5$, $\dot{r}_{0}=0.01$,
$\tau=1000$. More specifically, the left panel reports the GW frequency
(top row) as computed from the toy model and from an actual simulation
(marked by the index ``NR''), the corresponding GW amplitudes (middle
row) and the radial displacement $r(t)$ (bottom row). Note how the GW
frequencies and amplitudes are anti-correlated and the minimum in the
radial displacement at $t^{\psi_{4}}_{0}$ corresponds to the maximum
frequency, \ie $f^{\psi_{4}}_{0}$, and minimum amplitude, as expected in
a condition of quasi time-symmetry (in the toy model $f^{h}_{0}$ is
actually a local minimum of $f^h_{\rm GW}(t)$, a case often found in NR
data). The right panel of Fig.~\ref{fig:toy_model}, on the other hand, is
used to show the relation between the GW quantities of the toy model and
other dynamical quantities such as the time derivatives of the angular
velocity, $\partial_t\Omega(t)$ and $\partial^2_t\Omega(t)$, and of the
radial displacement, $\partial_tr(t)$ and $\partial^2_t r(t)$ (the top
row of the right panel contains the same information as in the top and
middle row of the left panel). It is clear that for the first few
oscillations the minima of the amplitude (maxima of the frequency)
correspond to moments in time when the angular frequency $\Omega$ attains
a local maximum, and the radial displacement a local minimum.
 
 We find it rather remarkable that the model originally developed to
 provide justification for the dominant postmerger frequency can be used
 to mimic the presence of the $f_{0}$ frequency with no additional
 adjustments, appearing as a result of chosen parameters and initial
 conditions. Finally, while we do not consider it here, it would be
 useful to generalise the model to admit unequal sphere masses, \ie
 $m/M<1$, and assess the impact it has on the $f_{0}$ frequency and on
 the evolution of the $m=2$ and $m=1$ modes.
 \begin{figure*}
   \centering 
   \includegraphics[height=0.04\textwidth]{Fig2_legend.pdf}\\
   \includegraphics[width=0.48\textwidth]{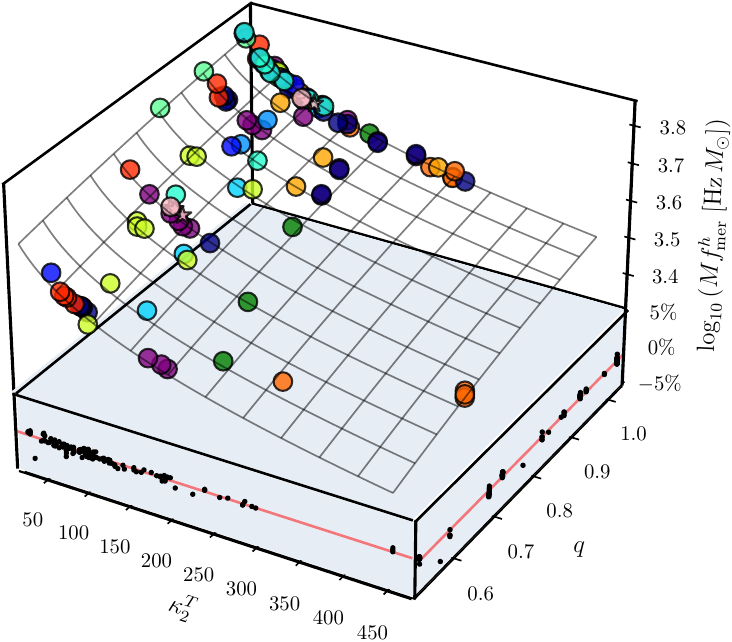}
   \hskip 0.2cm
   \includegraphics[width=0.48\textwidth]{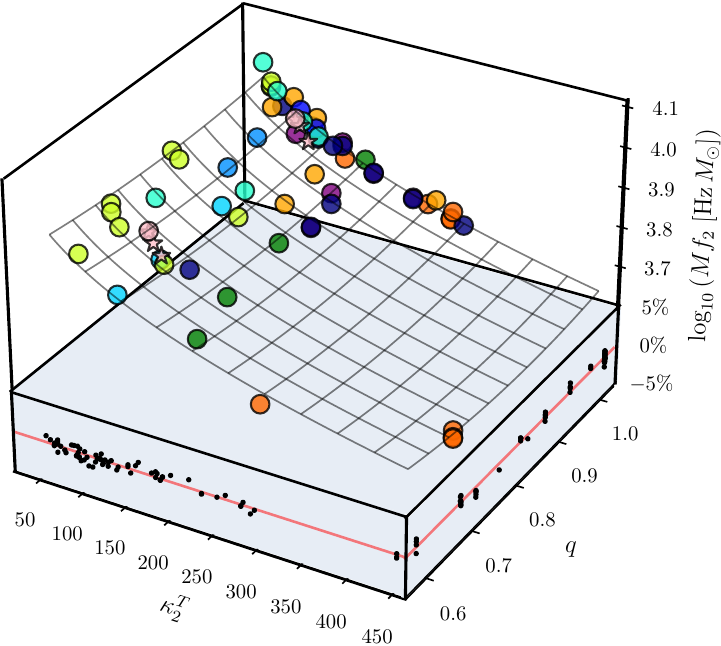}
   \caption{The same as in Fig.~$2$ in the main text but for the GW
     frequencies $f^{h}_{\rm mer}$ and $f_2$. Note how $\mathcal{F}_1$ is
     equally accurate in modelling other frequencies, also when obtained
     from the strain as in the case for $f^{h}_{\rm mer}$ and $f_2$.}
     \label{fig:2b}
 \end{figure*}
 
\subsection*{On the effectiveness of the fit}
 
To illustrate effectiveness of the fitting functional form
$\mathcal{F}_1$ in capturing the spectral properties of the signal in a
quasi-universal manner, we report in Fig.~$2$ the same 3D representation
as of the GW frequencies shown in Fig.~$2$ but for the GW frequencies
$f^{h}_{\rm mer}$ and $f_2$; note that $f^h_{\rm mer}$ is referred to as
$f_{\rm max}$ in~\cite{Takami2014, Rezzolla2016} and that the $f_{2}$
frequency corresponds to the peak of the PSD computed using the
definition in \cite{Tootle2022}, with the time integration encompassing
the full GW signal. Clearly, the fitting function $\mathcal{F}_{1}$ is
equally accurate in modelling other frequencies, quite independently of
whether they have been computed from the Weyl scalar or from the GW
strain, as it is as it is the case for $f^{h}_{\rm mer}$ and $f_2$.  We
also note that the equal-mass limit of the fitting function
$\mathcal{F}_1$ for the frequency $f^{h}_{\rm mer}$, recovers quite
accurately the simpler fit examined in~\cite{Takami2014} which expressed
the merger frequency as a first-order expression of the tidal
deformability $f^{h}_{\rm mer} = a_{0} + b
({\kappa_{2}^{_{T}}})^{{1}/{5}}$. When setting $q=1$ and $n={1}/{5}$ in
$\mathcal{F}_{1}$, we obtain $b=b_{0}+b_{1}+b_{2}=-0.199$ (\cf
Table~\ref{tab:fit_parameters}), thus resulting in a relative difference
that is $\lesssim2\%$ with respect to the value $b=-0.195$ found
in~\cite{Takami2014}. At the same time, we also note that the fit of
$f^{h}_{\rm mer}$ is overall slightly better than that for
$f^{\psi_4}_{\rm mer}$ (see Table~\ref{tab:fit_parameters}); while we do
not believe this to be statistically very significant, it may be due to
the slightly larger range in which $f^{\psi_{4}}_{\rm mer}$ is measured.

The only exception to the remarkably good fit of the frequencies is found
in the $f^{\psi_{4}}_{0}$ frequency, where a noticeable deviation from
the trend is visible for binary systems with equal or very-unequal
masses, regardless of the EOS employed (see lower part of the
bottom-right panel of Fig.~$2$). We attribute this decrease in accuracy
to the already described difficulties of measuring this frequency using
the $\psi_4$ GW signal, which are even more severe when using $|h|$. From
a physical point of view, the moment of time symmetry corresponds to the
time when the non-axisymmetric quadrupolar deformations of the HMNS are
at a minimum, which leads to a severely suppressed GW signal (indeed the
amplitude is at a minimum). This is particularly severe for binaries with
$q\simeq 1$, where the $\ell=2,m=2$ deformation is the largest and is
significantly suppressed at the moment of time-symmetry, and for binaries
with $q \ll 1$, where the $\ell=2,m=2$ deformation is further decreased
by the mass asymmetry. Notwithstanding these difficulties, relative
fitting error for $f^{\psi_{4}}_{0}$ [\ie $\max(\Delta f/f$)] is at most
$4.2\%$ and only $1.4\%$ on average [\ie $\langle|\Delta f|/f\rangle$].

\begin{table*}[t]
  \begin{ruledtabular}
    \begin{tabular}{lcccccccccc}
      Freq. & $a_{0}$ & $b_{0}$ & $b_{1}$ & $b_{2}$ & $n$ & $\langle \Delta f/f\rangle$ & $\max(\Delta f/f)$ & $\chi^{2}$ & $\chi^{2}_{\rm red}$ & $R^{2}$\\
            & & & & & & $[\%]$ & $[\%]$ & & $\times 10^{-3}$ & \\
      \hline
      $f^{\psi_{4}}_{\rm mer}$ & $4.589$ & $-0.581$ & $\phantom{-}0.543$ & $-0.236$ & $0.20\phantom{^{\dagger}}$  & $0.65$ & $4.52$ & $0.135$ & $1.19 $ & $0.934$  \\
      $f^{h}_{\rm mer}$      & $4.201$ & $-0.330$ & $\phantom{-}0.198$ & $-0.067$ & $0.20\phantom{^{\dagger}}$  & $0.32$ & $2.99$ & $0.035$ & $0.30 $ & $0.957$  \\
      $f^{\psi_{4}}_{\rm mer}$ & $6.067$ & $-2.142$ & $\phantom{-}0.970$ & $-0.410$ & $0.07^{\dagger}$            & $0.59$ & $4.24$ & $0.114$ & $1.00 $ & $0.938$  \\
      $f^{h}_{\rm mer}$      & $4.457$ & $-0.578$ & $\phantom{-}0.262$ & $-0.087$ & $0.14^{\dagger}$           & $0.32$ & $2.94$ & $0.034$ & $0.30 $ & $0.957$  \\
      $f^{\psi_{4}}_{0}$      & $6.550$ & $-2.099$ & $\phantom{-}0.518$ & $-0.304$ & $0.06^{\dagger}$            & $1.38$ & $4.22$ & $0.477$ & $4.92 $ & $0.611$  \\
      $f_{2}$               & $4.617$ & $-0.170$ & $-0.264$ & $\phantom{-}0.160$ & $0.20^{\dagger}$            & $0.45$ & $1.34$ & $0.030$  & $0.48 $ & $0.926$  \\
      \hline
    \end{tabular}
  \end{ruledtabular}
  \caption{Best-fit values for the coefficients of the functional form
    $\mathcal{F}_1$. Also reported are the maximal and average relative
    difference, as well as the $\chi^{2}$, $\chi^{2}_{\rm{red}}$ and
    $R^{2}$ coefficients of the fit.  Indicated with a $\dagger$ are the
    best-fit values of $n$ when this coefficient is constrained by the
    fit.}
        \label{tab:fit_parameters}
\end{table*}

\subsection*{Growth of the SNR ratio}

{In the main text we have discussed how to use the amplitude ratio
  $|\psi_{4}^{21}|/|\psi_{4}^{22}|$ as an effective proxy for the
  relevance of the $m=1$ mode deformation. We next demonstrate that a
  clear correspondence exists between finding
  $|\psi_{4}^{21}|/|\psi_{4}^{22}| \sim 1$ and the ratio of the SNRs in
  the two modes. We start by recalling that given the power spectral density
  of the $m$-th mode of the GW strain decomposition $\tilde{h}_m(f)$ 
  (such as the one presented in Fig.~\ref{fig:3D_PSD_amp_freq} for $m=2$) and a
  noise spectral density of the detector $S_{n}(f)$, the corresponding
  $m$-th SNR is defined as
}
  \begin{equation}
    {\rm SNR}_{m=k} := \left[\int_{0}^{\infty}
    4\frac{\vert \tilde{h}_{m=k}(f)\vert^2}{S_n(f)}df \right]^{\frac{1}{2}} \,.
    \label{eq:SNR}
  \end{equation}
Clearly, the ratio of the SNRs, ${\rm SNR}_{m=1}/{\rm SNR}_{m=2}$ and the
rate at which it evolves depends on the time $t_i$ when the signal starts
to be considered. Because there is no $m=1$ signal during the inspiral,
the SNR ratio would be intrinsically dominated by the $m=2$ component of
the signal if $t_i$ was chosen to be the time the signal entered the
detector. Hence, to fairly assess the growing contribution of the $m=1$
mode, we compute a time-windowed SNR over a running window of width
$\Delta T=5\,$ms. {We should remark that this approach is logically
  and mathematically equivalent to what is done when computing
  spectrograms~\citep[see, \eg][]{Abbott2017} and hence determines, at
  any given time, the characteristic frequency at which the GW is
  emitted. In essence,} for any time $\bar{t}$, we compute the SNR as
defined in Eq.~\eqref{eq:SNR}, where the signal in the time domain is in
the interval $t \in [\bar{t}-\Delta T/2, \bar{t}+\Delta T/2]$. This
time-windowed SNR provides an ``instantaneous'' measure of the ability of
a detector to measure a signal of given strength and is mathematically
equivalent to what is routinely done when computing spectrograms in GW
data analysis.

\begin{figure*}
  \centering 
  \includegraphics[height=0.5\textwidth]{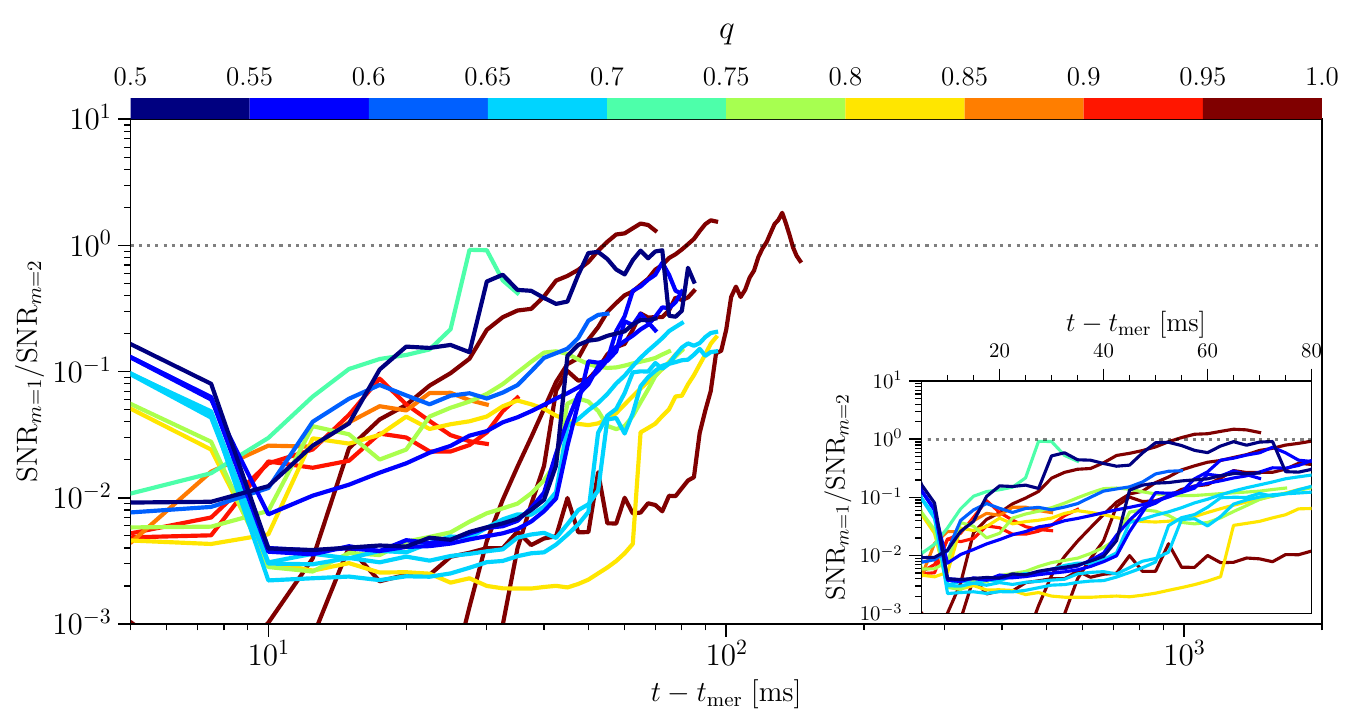}\\
  \caption{Instantaneous SNR ratio as computed for the selected set of
    configurations in~Fig.$4$ of the main text, where the time-window has
    been chosen to be $\Delta T =5\,$ms.  The detector's sensitivity is
    based on the O3 observing run of LIGO and the 
    angle of observation does not favour either
    of the modes. Note that ${\rm SNR}_{m=1}/{\rm SNR}_{m=2} \simeq
    \mathcal{O}(1)$ around the same time as
    $|\psi_{4}^{21}|/|\psi_{4}^{22}| \simeq \mathcal{O}(1)$.}
     \label{fig:snr_ratio}
\end{figure*}

As demonstrated in Fig.~\ref{fig:snr_ratio}, the SNR ratio computed in
this way grows to be of $\mathcal{O}(1)$ at the same time when
$|\psi_{4}^{21}|/|\psi_{4}^{22}| \simeq \mathcal{O}(1)$, thus supporting
the effectiveness of the mode ratio in acting as a proxy for the SNR
ratio. Finally, we note that the estimates provided in
Fig.~\ref{fig:snr_ratio} are similar in spirit to the SNR estimates
suggested by~\citet{Lehner2016a}. {We stress that the results shown
  should not be interpreted as pointing out to a \textit{global-in-time}
  dominance of the $m=1$ mode; rather, they suggest a an enhanced
  importance of this mode that is \textit{local-in-time} and appears only
  long past the merger. When considering the full GW signal, the $m=2$
  will always provide the largest contribution, by far.}

\end{document}